\documentstyle[11pt,paspconf,psfig]{article}

\begin{document}

\title{Joint formation of bright quasars and elliptical galaxies in
the young Universe}

\author{Pierluigi Monaco}
\affil{Dipartimento di Astronomia, Universit\`a di Trieste, and SISSA, Trieste}

\author{Paolo Salucci and Luigi Danese}
\affil{SISSA, Trieste}

\begin{abstract}
We show that the mass function of black holes expected from the past
quasar activity (both visible and obscured) is consistent with the
number of dormant black holes found in the bulges of nearby galaxies.
The joint formation of quasars and bulges is addressed by means of an
analytical model for galaxy formation, based on the hierarchical
clustering of cold dark matter halos. The model is able to reproduce
the main statistical properties of both populations under the
hypotheses that (i) star formation and quasar shining follow an
anti-hierarchical order, and (ii) galaxy morphology and final black
hole mass are determined by the same physical process.
\end{abstract}


\section{Introduction}

Quasars, and AGN in general, are often supposed to be powered by
accretion of gas into supermassive black holes (BHs).  In this case,
large dormant BHs are expected in the nuclei of nearby galaxies
(Soltan 1982; Rees 1984; Cavaliere \& Padovani 1986).  Assuming
accretion at a known fraction of the Eddington rate and efficiency of
radiation of 10\% in units of $mc^2$, it is possible to estimate the
expected mass function of dormant BHs.  This mass function implies a
number density of large BHs ($>10^8$ M$_\odot$) compatible with the
hypothesis that a BH is present in each bright bulge.

Recent detailed observations of the cores of nearby galaxies have lead
to the discovery of massive dark objects in most cases (Magorrian et
al. 1998; van der Marel 1999).  Even though many details are still
uncertain, many authors agree in claiming a correlation between the
mass of the massive dark object and the host bulge.  Interpreting
these dark objects as the expected dormant BHs, the BH -- bulge
correlation strongly suggests a connection between quasar activity and
the formation of galactic bulges.

\section{The mass function of dormant black holes.}

It is assumed that the accretion of matter onto a BH of mass
$M_\bullet$ is a fixed fraction $f_{\rm ED}$ of the Eddington rate, so
that the quasar luminosity is $L=f_{\rm ED} L_{\rm ED}$, where $L_{\rm
ED} \simeq 3.4\ 10^{4} (M_\bullet/M_\odot)\ L_\odot$.  The efficiency
of accretion $f_{\rm ED}$ is assumed to increase from 0.1 for the
smallest BHs ($\sim 10^6$ M$_\odot$) to 1 for the largest ones ($\sim
10^{10}$ M$_\odot$).  Then, the mass function of dormant BHs is
calculated by integrating the luminosity function of quasars (see
Salucci et al. 1999a for details).  We assume that a significant
fraction of AGNs are heavily obscured, and give the dominant
contribution to the hard X-ray cosmological background (see, e.g.,
Celotti et al. 1995; Comastri et al. 1995; Fiore et al. 1998).  We
include such objects using the model of Comastri et al.  (1995).  The
resulting expected mass function of dormant BHs is shown in Fig. 1
(dashed line), for $H_0=70$ km/s/Mpc and $\Omega=1$.

\begin{figure}
\psfig{figure=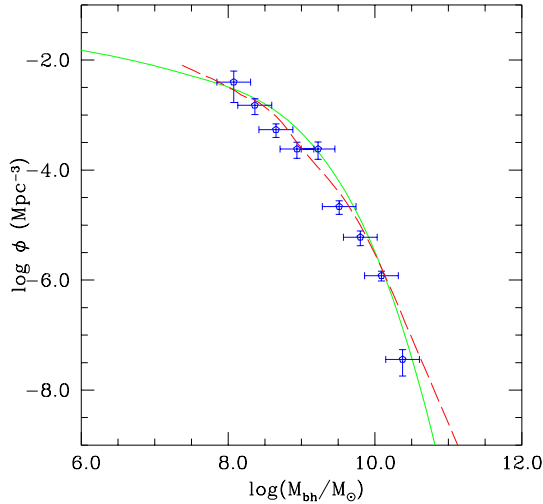,width=8cm}
\caption{Mass function of dormant BHs (see text for details).}
\end{figure}

The mass function of dormant BHs residing in nearby galaxies is
estimated with two independent methods (see Salucci et al. 1999a for
details).  Firstly, the mass function of galactic bulges is convolved
with a fiducial BH -- bulge relation (a lognormal, with width 0.3 dex
and average $M_\bullet/M_{\rm bulge}\sim 3\ 10^{-3}$).  Fig. 1 shows
the resulting mass function (continuous line).  Secondly, exploiting
the correlation between radio power from the core of elliptical
galaxies and BH mass ($P \propto M_\bullet^\alpha$, where $\alpha\sim
2\ -\ 2.2$, see also Franceschini et al. 1998), the radio luminosity
function of elliptical cores is convolved with a BH -- radio power
relation to obtain another estimate of the mass function of the
dormant objects.  The result is again shown in Fig. 1 (points with
errorbars).

\section{A favoured scenario}

The three determinations of the mass function of dormant BHs agree for
reasonable values of the parameters involved.  This highlights a
dichotomy (in a statistical sense) in the behaviour of BHs.  Larger
objects ($M>10^8$ M$_\odot$) are hosted in ellipticals, shine as
bright quasars at high redshift, almost at the Eddington luminosity,
are hardly reactivated and hardly obscured, while smaller BHs
($M<10^8$ M$_\odot$) are hosted in the bulges of spiral galaxies,
shine also at low redshift with a lower luminosity (in Eddington
units), and may be reactivated and obscured.

The abundance of BHs in the bulges of spiral galaxies is more
difficult to estimate.  Upper limits have been determined by Salucci
et al. (1999b) by analyzing nearly a thousand rotation curves for
spirals.

\begin{figure}
\psfig{figure=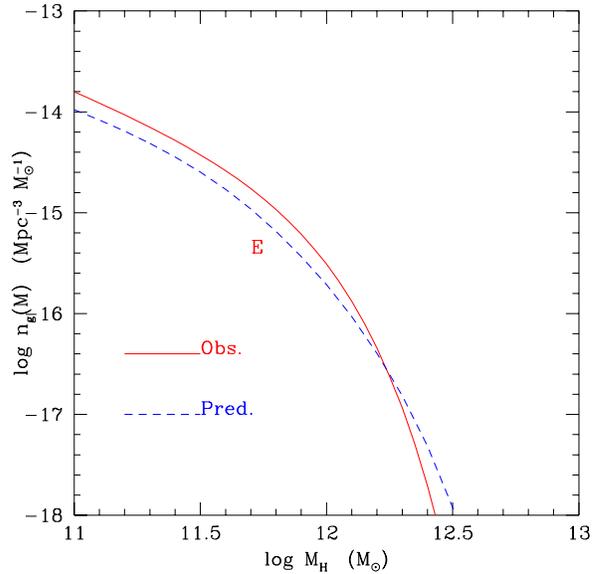,width=8cm}
\caption{Mass function of dark matter halos of ellipticals.}
\end{figure}

\section{A model for the joint formation of quasars and 
elliptical galaxies}

We have constructed an analytical model for the joint formation of
ellipticals and quasars in the framework of hierarchical CDM models.  The
details are given in Monaco, Salucci \& Danese (1999).

In a bulge, quasar activity and the main burst of star formation,
which mark the main ``shining phase'' of a galactic dark matter halo,
are likely to be close in time (see, e.g., Hamann \& Ferland 1993; see
also Best, these proceedings).  It is supposed that the shining phase
of a galactic halo is delayed with respect to its dynamical formation.
This delay is assumed to be small for the halos corresponding to large
ellipticals, and increasingly larger for smaller halos.  In this way
the hierarchical order is inverted for halo shining.  This is done to
reproduce the apparent anti-hierarchical evolution of quasars while
preserving a correlation between bulge and BH mass.

\begin{figure}
\psfig{figure=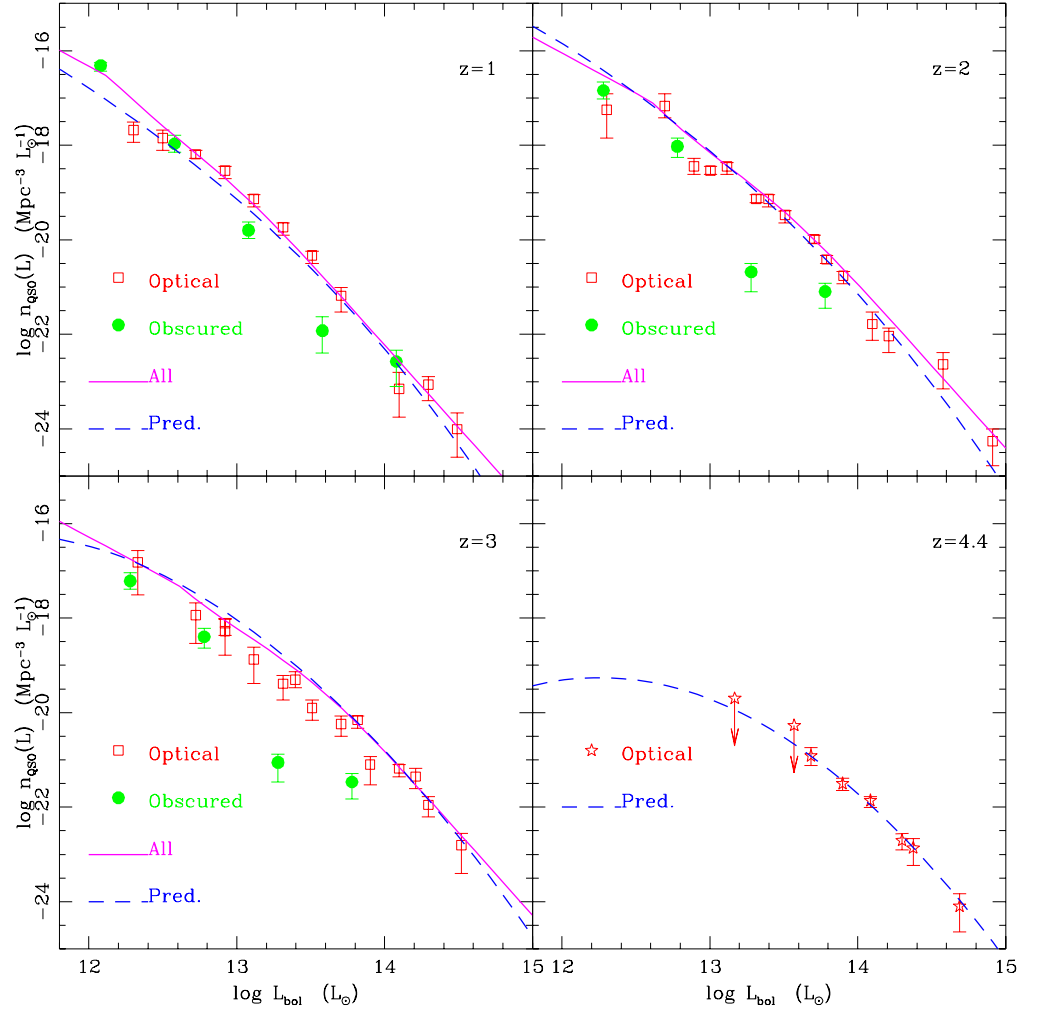,width=15cm}
\caption{Quasar luminosity functions at various redshifts ($H_0 = 70$
km/s/Mpc, $\Omega=0.3$, $\Omega_\Lambda=0.7$). The optical luminosity
function is taken from Pei (1995). Following Comastri et al. (1995),
the contribution of obscured objects is estimated from the X-ray
luminosity function of Boyle et al. (1993).  The curve named ``ALL''
gives the total contribution of optical and obscured objects. At
$z=4.4$ the data points of Kennefick et al.  (1996) (denoted by stars)
are reported.}
\end{figure}

The mass of the BH formed during the shining phase is assumed to
depend on the halo mass, and to be modulated by the same variable
which determines the morphological type, so as to obtain a BH-bulge
relation.  This variable is assumed to be either the spin of the dark
matter halo or the fraction of the merging masses at the formation
time.

The model reproduces successfully the main observable quantities
relative both to elliptical galaxies and quasars; see Monaco et al.
(1999) for details.  The results shown in this paper are for a flat
CDM model with $\Omega=0.3$, cosmological constant and $H_0=70$
km/s/Mpc.  Fig. 2 shows the predicted mass function for the dark
matter halos of ellipticals, compared with that inferred from the
luminosity function and reasonable hypotheses on the mass-to-light
ratios of ellipticals.  Fig. 3 shows the comparison between the
predicted and observed quasar luminosity functions at different
redshifts.  The data are taken from Pei (1995), Boyle et al. (1993)
($z=1$ to 3) and Kennefick, Djorgovski and Meylan (1996) ($z=4.4$).

\acknowledgments

The authors thank E. Szuszkiewicz and C. Ratnam for discussions.

\end{document}